\documentclass[conference,a4paper]{IEEEtran}
\usepackage{cite}
\usepackage{amsmath,amssymb,amsfonts}
\usepackage{graphicx}
\usepackage{wasysym}
\usepackage{subcaption}
\usepackage{multicol}
\usepackage{tabularx}
\usepackage{bm}
\usepackage[noend]{algpseudocode}
\usepackage{algorithmicx,algorithm}
\usepackage{diagbox}

\usepackage{array}
\usepackage{fixltx2e}
\usepackage{dblfloatfix}
\usepackage{url}
\usepackage{newclude}
\usepackage{comment}
\usepackage[dvipsnames]{xcolor}
\usepackage{fancyhdr}
\begin{document}

%
\title{Features-over-the-Air: Contrastive Learning Enabled Cooperative Edge Inference}


\author{%
   \IEEEauthorblockN{Haotian Wu,~\IEEEmembership{Graduate Student Member,~IEEE,}
                    Nitish Mital,~\IEEEmembership{Member,~IEEE,}
Krystian Mikolajczyk,~\IEEEmembership{Senior Member,~IEEE,}
Deniz G\"{u}nd\"{u}z ~\IEEEmembership{Fellow,~IEEE}}
   
   \IEEEauthorblockA{Department of Electrical and Electronic Engineering, Imperial College London, London SW7 2BT, UK\\
    Email:$\left\{haotian.wu17, n.mital, k.mikolajczyk, d.gunduz\right\}$ @imperial.ac.uk}

   
}

\IEEEtitleabstractindextext{%
\begin{abstract}
We study the collaborative image retrieval problem at the wireless edge, where multiple edge devices capture images of the same object, which are then used jointly to retrieve similar images at the edge server over a shared multiple access channel. We propose a semantic non-orthogonal multiple access (NOMA) communication paradigm, in which extracted features from each device are mapped directly to channel inputs, which are then added over-the-air. 
We propose a novel contrastive learning (CL)-based semantic communication (CL-SC) paradigm, aiming to exploit signal correlations to maximize the retrieval accuracy under a total bandwidth constraints. Specifically, we treat noisy correlated signals as different augmentations of a common identity, and propose a cross-view CL algorithm to optimize the correlated signals in a coarse-to-fine fashion to improve retrieval accuracy. Extensive numerical experiments verify that our method achieves the state-of-the-art performance and can significantly improve retrieval accuracy, with particularly significant gains in low signla-to-noise ratio (SNR) and limited bandwidth regimes.



 
\end{abstract}

}

\maketitle
\thispagestyle{fancy} 
\cfoot{\thepage}
\renewcommand{\headrulewidth}{0pt} 
\renewcommand{\footrulewidth}{0pt} 
\pagestyle{fancy}
\cfoot{\thepage}

\IEEEdisplaynontitleabstractindextext

\section{Introduction}
\IEEEPARstart{T}he recent success of machine learning algorithms has triggered significant interest in developing semantic communication systems\cite{xie2021deep,gunduz2020communicate, gunduz2022beyond}, where goal-oriented semantic content of signals are taken into account when designing communication schemes, especially in the distributed edge inference problems \cite{jscc_ae, shao2021learning, jscc_ae_4, just_draft}. 

In distributed inference problems, deep neural network models are often employed across multiple distributed devices with limited communication resources, where data must be communicated between nodes to increase inference accuracy. In particular, in a collaborative image retrieval task, the edge devices try to identify the same object in a gallery database from the images taken by different cameras. Unlike other edge classification or inference problems \cite{shao2021learning,jscc_ae_4}, which can be carried out locally at the edge device with enough computational resources, remote inference is essential for the image retrieval task even with a single device, as the gallery database is only available at the edge server. 

\begin{figure}[!t]
\centering
\includegraphics[width=0.9\linewidth]{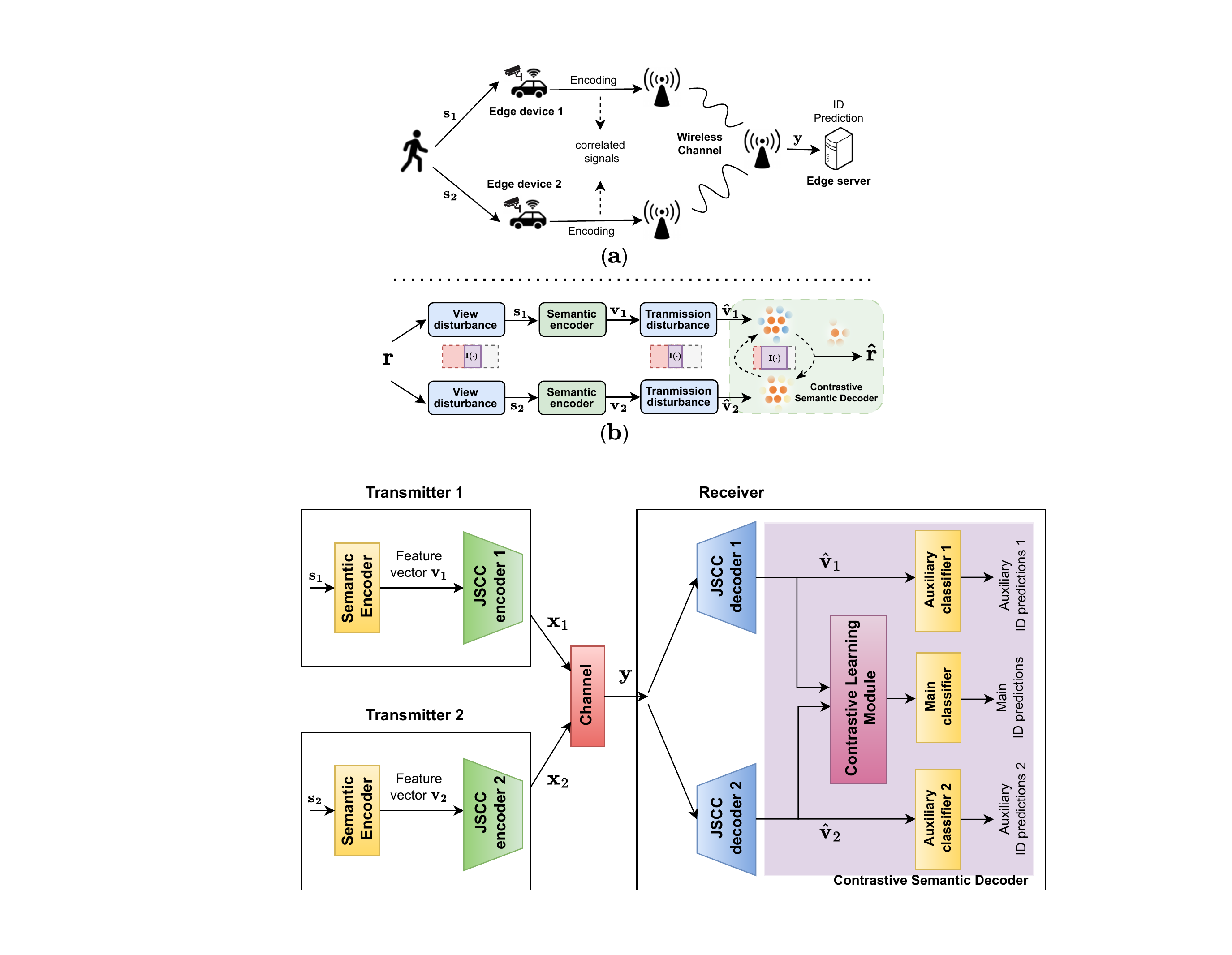}
\caption{(a) Two-source collaborative remote image retrieval problem. (b) The role of CL in this problem, where the dotted and solid rectangles represent the information contained in each observation and the shared identity-related information. The common identity-related features are represented by the orange points, while the inconsistent features with other colors. We expect CL module to help $\bm{\hat{v}_1}$ and $\bm{\hat{v}_2}$ to be more discriminative with maximal semantic consistency and identity-related features.}
\label{fig:person_reid}
\end{figure}

Fig. \ref{fig:person_reid}(a) illustrates a typical collaborative remote inference problem, where two edge devices cooperate to perform the inference task over a shared multiple access channel (MAC). Considering the communication latency, bandwidth, and power constraints, only the most relevant semantic features must be extracted at the edge devices and transmitted to the edge server \cite{jscc_ae,shao2021learning}. This calls for semantic communication, as we extract and convey the most relevant features to represent the semantic content of the source image for transmission  \cite{gunduz2022beyond}.

In a conventional communication system, these features are extracted at the application layer and are conveyed to the receiver using existing channel coding and modulation techniques. However, the latency requirements of many edge applications result in extremely low short blocklength codes, and the suboptimality of the separate source-channel coding schemes increases as the code length diminishes. An autoencoder-based joint source and channel coding (JSCC) wireless retrieval scheme was pioneered in \cite{jscc_ae}, which is shown to outperform the separation-based digital approach by a large margin under all channel conditions. When it comes to multiple networks, however, even the theoretical optimality of separation between the source and channel coding ceases to hold in general \cite{shannon1961two, 1056273}. The remote retrieval problem with multiple users is also studied in \cite{just_draft}, where a collaborative semantic communication scheme is proposed. It is shown that, given the same amount of communication resources, inference accuracy can be improved by incorporating extra information from multiple views and employing deep learning-based JSCC for the relevant feature transmission. For the transmission of features, \cite{just_draft} studies both an orthogonal multiple access (OMA) scheme, in which the two edge devices transmit on orthogonal channels (e.g., time-division multiple access); and (2) non-orthogonal multiple access channel (NOMA) scheme, in which both edge devices transmit over the shared channel. It is shown in \cite{just_draft} that NOMA wireless transmission scheme provides significant gains by preserving the correlations between the inputs in the transmitted symbols over the channel. Inspired by this result, here we also focus on NOMA transmission.

In the multi-terminal edge retrieval problem under consideration, the distributed nature poses unique collaborative challenges, requiring the edge devices to collaboratively extract the relevant semantic information while discarding unnecessary information to improve inference accuracy. 
In order to learn a discriminative representation of the common semantics between correlated inputs, contrastive learning (CL), which explores the discriminative features from different data augmentations, is a promising method. The core idea of CL is to learn contrastive features by minimizing the feature distance between samples with the same identity, while maximizing the feature distance between samples with different identities. Typical CL methods, such as SimCLR\cite{chen2020simple} and BYOL\cite{grill2020bootstrap}, show competitive performance in various tasks. The CL method can improve the representation learning process by discriminating semantics, sharing the same objective of the edge inference task in the emerging semantic communication system, where contrastive semantic features can provide significant benefits for edge inference, such as pre-processing and disentangling the raw data \cite{chaccour2022disentangling}.
 
Therefore, in order to improve the inference performance, we propose a CL-based collaborative semantic communication scheme, which optimizes the extracted features in a coarse-to-fine fashion based on their correlations. Specifically, we model the problem as a multi-terminal and noisy channel variant of the remote source compression problem \cite{1057738}, as shown in Fig. \ref{fig:person_reid} (b), where we model the source images $\bm{s}_1$ and $\bm{s}_2$ from different cameras as the distinct noisy versions of the desired identity $\bm{r}$ through stochastic channels $p(\bm{s}_1 \vert \bm{r})$ and $p(\bm{s}_2 \vert \bm{r})$, where the varying backgrounds and viewing angles can introduce extraneous information. Following previous pipelines \cite{jscc_ae, just_draft}, we encode the source signals into semantic features $\bm{v}_1$ and $\bm{v}_2$, which are transmitted over the MAC and then reconstructed as the coarse semantic features by a JSCC scheme. This process is modeled as the `transmission disturbance' $p(\hat{\bm{v}}_1\vert \bm{v}_1)$ and $p(\hat{\bm{v}}_2\vert \bm{v}_2)$, aiming to maximize the reference accuracy instead of minimizing the transmission distortion. The reconstructed features are then input to a contrastive semantic decoder to output the inference identity $\hat{\bm{r}}$, where a cross-view CL method is proposed to refine the coarse semantic features by maximizing the common identity-related information between different disturbed views with a common identity. 

Our main contributions can be summarized as follows:

\begin{itemize}
\item We propose a novel CL paradigm for edge inference. We study the effect of the channel condition on the properties of semantic information transmitted by different edge devices. The proposed cross-view CL paradigm is easy to interpret and can be used in any scenario where multiple correlated signals exist.

\item The main novelty of this paper is the use of CL for collaborative inference among different wireless edge devices. Numerical experiments verify that our method improves upon the state-of-the-art in all the considered channel conditions. Exploiting CL technologies can significantly improve retrieval performance, especially in the low signal-to-noise ratio (SNR) and limited bandwidth regimes.
\end{itemize}

\section{System model}
We consider two distinct devices acquiring images of the same identity from different angles and potentially at different qualities. They communicate over a shared MAC with an edge server, where inference is performed to retrieve the image of the same identity in a local database.

We denote the image observed by the $i$-th transmitter as $\bm{s}_i \in \mathbb{R}^{n}$, $i=1,2$. The $i$-th transmitter encodes the $\bm{s}_i$ into a complex channel codeword $\bm{x_i}=\mathcal{E}_i(\bm{s_i}): \mathbb{R}^{p} \rightarrow \mathbb{C}^{k}$, where $k$ represents the allocated channel bandwidth. 
The channel input $\bm{x}_i$ of each transmitter is subject to the power constraint $\frac{1}{k}\vert\vert \bm{x}_i \vert\vert^2_2 \leq 1$.

In the NOMA transmission scheme, encoded channel symbols $\bm{x}_1$ and $\bm{x}_2$ are transmitted simultaneously over the shared wireless channel. 
The channel output is given by $\bm{y}=h_1\bm{x_1}+h_2\bm{x_2}+\bm{w}$, where $\bm{w}\in \mathbb{C}^{k}$ is the additive white Gaussian noise (AWGN) term with independent and identically distributed (i.i.d.) samples from a complex Gaussian distribution $\mathcal{CN}(0,\sigma_w^2)$. For an AWGN channel, the channel gains are fixed as $h_1=h_2=1$. For the slow fading channel, the channel gains are sampled from a complex Gaussian distribution as $h_1, h_2 \sim \mathbb{C}\mathcal{N}(0, 1)$, and remain constant for $k$ channel uses. The channel quality is measured by the average channel SNR, defined as: $\mu\triangleq 10\log_{10}\frac{1}{\sigma_w^2}$ dB. The receiver performs image retrieval using $\bm{y}$, and our goal is to maximize the retrieval accuracy despite the presence of noise and fading over the channel, and to identify how the devices should exploit the shared channel resources. 

\begin{figure}[!t]
\begin{center}
   \includegraphics[width=0.9\linewidth]{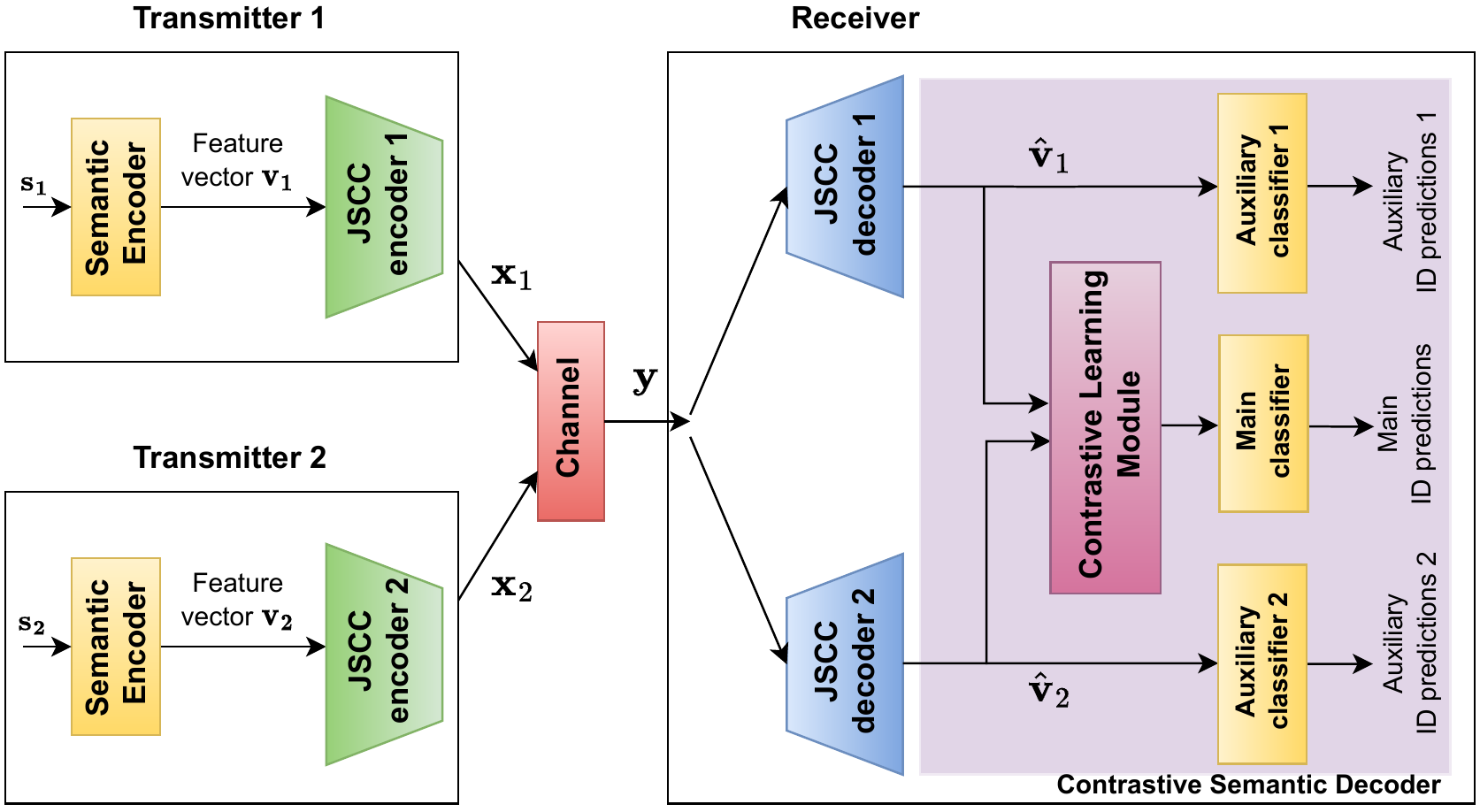}
   \end{center}
   \caption{The pipeline of the proposed CL-SC scheme.
   } 
\label{pipeline}
\end{figure}

\begin{figure*}[!t]
\begin{center}
   \includegraphics[width=0.69\linewidth]{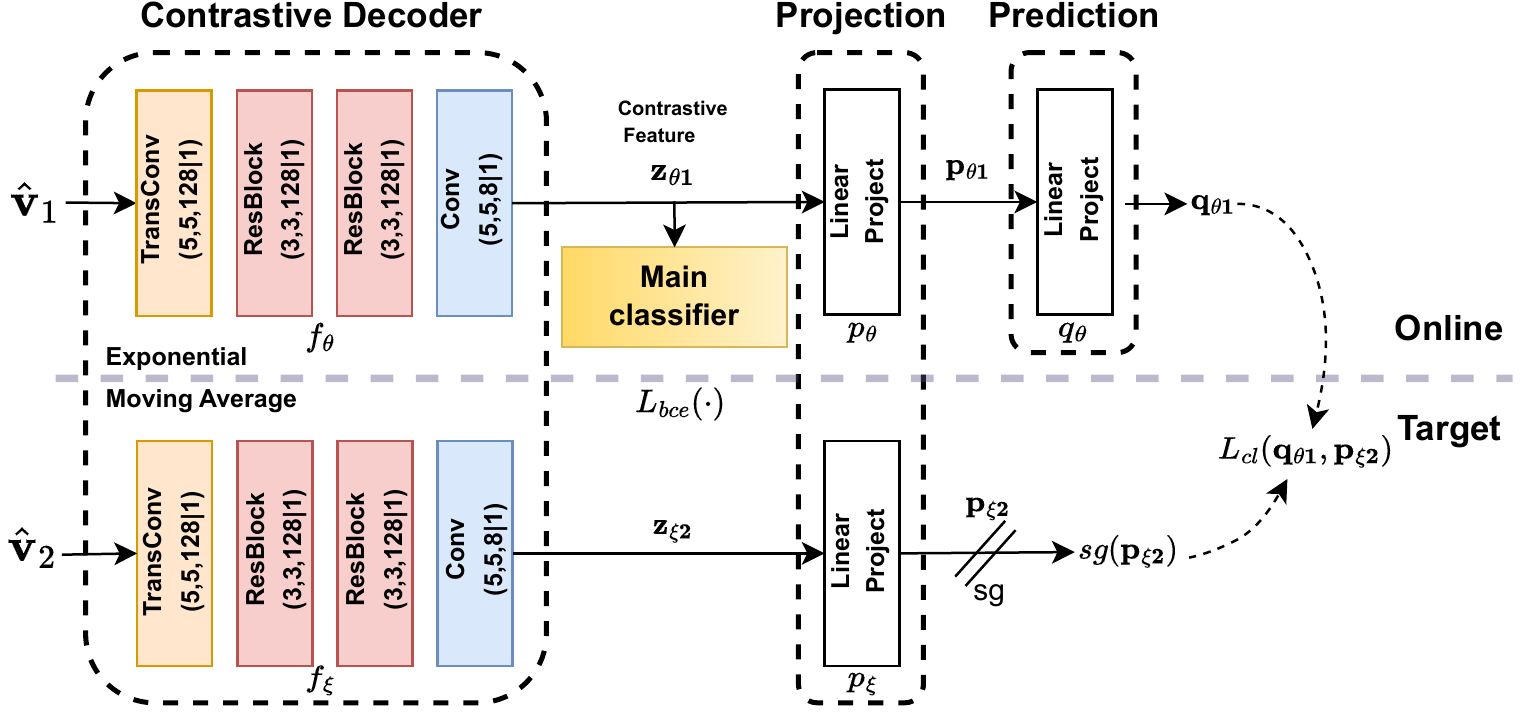}
   \end{center}
   \caption{Block diagrams of the CL module, where each block consists of convolutional/deconvolutional operations with the format of $(\text{kernel size},\text{kernel size}, \text{out channel}|\text{stride})$ followed by batch normalization and ReLU activation. The Linear Project module consists of two linear layers followed by batch normalization and ReLU activation. `sg' denotes the stop-gradient operation. Feature vectors $\bm{\hat{v}_1}$ and $\bm{\hat{v}_2}$ are fed into $f_{\bm{\theta}}$ and $f_{\bm{\xi}}$, respectively, which are then exchanged with each other. }
\label{CL_model}
\end{figure*}


\section{Proposed Method}
We propose a CL based semantic communication scheme (CL-SC), illustrated in Fig. \ref{pipeline}. In the proposed paradigm, each transmitter consists of a semantic feature extractor and a JSCC encoder. The receiver comprises of symmetrical JSCC decoders and a CL module. 

\subsection{Retrieval baseline}
Considering the state-of-the-art and for fair comparisons, we apply the same retrieval baseline used in \cite{jscc_ae,just_draft}. Each transmitter employs a ResNet50 as the semantic encoder on the source signal $\bm{s_{i}}$, resulting in a $2048$-dimensional semantic vector $\bm{v_{i}}$, which is then fed into the JSCC encoder. In our implementation, receiver first decodes the received signal to recover the feature vectors from both transmitters as $\hat{\bm{v}}_i=\mathcal{D}_i(\bm{y}): \mathbb{C}^{k} \rightarrow \mathbb{R}^{p}$, $i=1,2$. Then it performs the retrieval task to find the identity $\bm{r}$ from the database based on $\hat{\bm{v}}_1, \hat{\bm{v}}_2$.

Then, the server computes the similarities between the viewpoints from the database gallery and each query image. A nearest neighbor search is performed over the concatenated similarity vector to retrieve the most likely candidate in the gallery. Aligned with previous works, during the training phase, we introduce two auxiliary classifiers to prevent vanishing gradients and improve the performance. 
Specifically, auxiliary classifiers are first used to pre-train the semantic encoders, and then to help jointly train the whole pipeline. In the inference phase, we discard all the classifiers and directly use the CL module's optimized features for image retrieval. Below we will explain the CL module in detail. 

\begin{algorithm}[t] 
\caption{Cross-view CL at the Edge}
\textbf{Input:} \\
Reconstructed semantic feature vectors: $\bm{\hat{v}_1}$, $\bm{\hat{v}_2}$\\
One-hot identity label vector and main classifier: $\bm{r}$, $I_{\bm{\theta}}$\\
Online decoder, projector, predictor: $f_{\bm{\theta}}$, $p_{\bm{\theta}}$, $q_{\bm{\theta}}$\\ 
Target decoder, projector:$f_{\bm{\xi}}$, $p_{\bm{\xi}}$\\
 Optimizers, target decay rate: $Opt_1$, $Opt_2$, $\tau$\\
\textbf{Output:} Optimized contrastive features: $\bm{z_{\theta_1}}$,$\bm{z_{\theta_2}}$\\
\rule{\linewidth}{0.05em}
\textbf{Training phase:} \\
\vspace{-13pt}
\begin{algorithmic}[1]
  \For{each epoch}
    \State \texttt{$\bm{z_{\theta_1}}=f_{\bm{\theta}}(\bm{\hat{v}_1})$, $\bm{z_{\theta_2}}=f_{\bm{\theta}}(\bm{\hat{v}_2})$} \Comment{ Online encoding}
    \State \texttt{$\mathcal{L}_{{ce}}=\textit{CE}(I_{\bm{\theta}}(\bm{z_{\bm{\theta}_1}}),\bm{r})+\textit{CE}(I_{\bm{\theta}}(\bm{z_{\bm{\theta}_2}})),\bm{r})$}

    \State \texttt{$\bm{\theta} \gets Opt_1(\bm{\theta},\nabla_{\bm{\theta}} \mathcal{L}_{{ce}})
$}
\Comment{Update online network}
    \vspace{+3pt}
    \If{epoch$\% 2=0$} 
    \State \texttt{$\bm{z_{\xi_1}}=f_{\bm{\xi}}(\bm{\hat{v}_1})$, $\bm{z_{\xi_2}}=f_{\bm{\xi}}(\bm{\hat{v}_2})$} \Comment{Target encoding}

    \State \texttt{$\bm{p_{\theta_1}}= p_{\bm{\theta}}(\bm{z_{\theta_1}})$}, \texttt{$\bm{p_{\theta_2}}= p_{\bm{\theta}}(\bm{z_{\theta_2}})$} 
    \Comment{Online project}
    \State \texttt{$\bm{p_{\xi_1}}= p_{\bm{\xi}}(\bm{z_{\xi_1}})$}, \texttt{$\bm{p_{\xi_2}}= p_{\bm{\xi}}(\bm{z_{\xi_2}})$} 
    \Comment{Target project}

        \State \texttt{$\bm{q_{\theta_1}}= q_{\bm{\theta}}(\bm{p_{\theta_1}})$}, \texttt{$\bm{q_{\theta_2}}= q_{\bm{\theta}}(\bm{p_{\theta_2}})$} 
        \Comment{Online predict}
    \State \texttt{$\mathcal{L}_{cl}=4-2(\frac{<\bm{q_{\theta_1}},\bm{p_{\xi_2}}>}{\lVert \bm{q_{\theta 1}} \rVert_2 \lVert \bm{p_{\xi_2}}\rVert_2}+\frac{<\bm{q_{\theta_2}},\bm{p_{\xi 1}}>}{\lVert \bm{q_{\theta_2}} \rVert_2 \lVert \bm{p_{\xi_1}}\rVert_2})$}
    \State \texttt{$\bm{\theta} \gets Opt_2(\bm{\theta},\nabla_{\bm{\theta}} \mathcal{L}_c)
$} \Comment{Update online network}
     \State \texttt{$\bm{\xi} \gets \tau\bm{\xi} +(1-\tau)\bm{\theta}$} \Comment{Update target network}
     \EndIf 
  \EndFor
\end{algorithmic}
\vspace{-10pt}
\rule{\linewidth}{0.05em}
\textbf{Inference phase:} \\
\vspace{-13pt}
\begin{algorithmic}[1]
    \State \texttt{$\bm{z_{\theta_1}}=f_{\bm{\theta}}(\bm{\hat{v}_1})$, $\bm{z_{\theta_2}}=f_{\bm{\theta}}(\bm{\hat{v}_2})$} \Comment{Optimized feature}
 \end{algorithmic}
\label{alg:CL}
\end{algorithm}


Inspired by the recent literature showing its superiority for remote inference and signal recovery problems \cite{jscc_ae, just_draft, wu2022vision,wu2022channel}, we employ JSCC with a similar structure to those used in \cite{jscc_ae, just_draft}. 
We jointly train the JSCC encoders $\mathcal{E}_1,\mathcal{E}_2$ and decoders $\mathcal{D}_1, \mathcal{D}_2$ by minimizing the mean squared error (MSE) loss function as: $\mathcal{L}_{mse}=\sum_{i=1}^{2} \mathbb{E}\big[ \|\bm{v_i}-\bm{\hat{v}_i}\|^2_2\big],$ where the expectation is taken over the randomness both in the source and channel distributions.

\subsection{CL module}
We propose a cross-view CL strategy to optimize the received semantic features. The architecture and operations of our CL module are elaborated in Fig. \ref{CL_model} and Algorithm \ref{alg:CL}. Inspired by\cite{chen2020simple,grill2020bootstrap,lin2021completer}, where it has been shown that cross-view prediction strategy can optimize the feature representations by maximizing the mutual information and minimizing the conditional entropy between different augmentations or views, our CL module employs a cross-view prediction strategy to optimize $\bm{\hat{v}_1}$ and $\bm{\hat{v}_2}$ for more representative semantics. Specifically, we treat $\bm{\hat{v}_1}$ and $\bm{\hat{v}_2}$ as different noisy observations of the source identity, as in remote source coding problems \cite{1057738}, and minimize the conditional entropy between them by cross-prediction operations. Our CL module consists of an online network and a target network, where the online network with input $\bm{\hat{v}_i}$ is optimized to predict the regression targets generated from the target network with input  $\bm{\hat{v}_j}$, and the target network is optimized by an average moving strategy to avoid network collapse. 

As shown in Fig. \ref{CL_model}, online and target networks share the same structure, but with different parameters $\bm{\theta}$ and $\bm{\xi}$. Both networks consist of a contrastive decoder $f_{\bm{\theta}}$ and $f_{\bm{\xi}}$, and a projection layer $p_{\bm{\theta}}$ and $p_{\bm{\xi}}$, respectively. The online network has an additional prediction layer $q_{\bm{\theta}}$. We represent the operations of the contrastive decoders in the online network and target network over the input $\bm{\hat{v}_i}$ as $\bm{z_{\theta_i}}=f_{\bm{\theta}}(\bm{\hat{v}_i})$ and $\bm{z_{\xi_i}}=f_{\bm{\xi}}(\bm{\hat{v}_i})$, respectively. By performing the cross-view prediction training, the online contrastive decoder is expected to optimize $\bm{\hat{v}_i}$ with more representative semantics as $\bm{z_{\theta_i}}$. Inspired by \cite{chen2020simple, grill2020bootstrap}, we empirically design an additional projection layer before the prediction operation for better performance. The operations of the projection layer for the online and target networks are denoted as $\bm{p_{\theta i}}=p_{\bm{\theta}}(\bm{z_{\theta_i}})$ and $\bm{p_{\xi i}}=p_{\bm{\xi}}(\bm{z_{\xi_i}})$, respectively. The output of the prediction layer is denoted as $\bm{q_{\theta i}}=q_{\bm{\theta}}(\bm{\bm{p_{\theta i}}})$. Considering the bi-view input data, each view $\bm{\hat{v}_i}$ will be fed into the online network as $\bm{q_{\theta_i}}$ to predict the regression target $\bm{p_{\xi_j}}$ from the target network with another view $\bm{\hat{v}_j}$. 

To train the CL module, we introduce a classification loss $\mathcal{L}_{{ce}}$ and a cross-view prediction loss $\mathcal{L}_{cl}$, where we optimize $\mathcal{L}_{ce}$ every epoch and $\mathcal{L}_{cl}$ every two epochs. Specifically, we introduce a `main classifier' to train $\mathcal{L}_{{ce}}$, which is a fully-connected layer followed by a softmax operation, denoted as $I_{\bm{\theta}}$. We define $\mathcal{L}_{{ce}}\triangleq \frac{1}{2} (\textit{CE}(I_{\bm{\theta}}(\bm{z_{\bm{\theta}_i}}),\bm{r})+\textit{CE}(I_{\bm{\theta}}(\bm{z_{\bm{\theta}_j}}),\bm{r})),$ where $\bm{r}$ is the one-hot identity label vector, $\textit{CE} (\cdot)$ is the cross entropy loss. Note that this main classifier is only introduced in the training phase and will be removed in the inference phase, where $\bm{z_{\theta_i}}$ and $\bm{z_{\theta_j}}$ are directly used for retrieval. The online network parameters $\bm{\theta}$ are then updated by the optimizer $Opt_1$ as: $\bm{\theta} \gets Opt_1(\bm{\theta},\nabla_{\bm{\theta}} \mathcal{L}_{{ce}})$ over $\mathcal{L}_{{ce}}$. $\mathcal{L}_{cl}$ measures the distance between the normalized online prediction output $\bm{\bar{q}_{\theta_i}}\triangleq\frac{\bm{q_{\theta_i}}}{\|\bm{q_{\theta_i}}\|_2}$ and the normalized target projection output $\bm{\bar{p}_{\xi_i}}\triangleq\frac{\bm{p_{\xi_i}}}{\|\bm{p_{\xi_i}}\|_2}$ as:
\begin{equation}
\begin{split}
\mathcal{L}_{cl}=&\|\bm{\bar{q}_{{\theta}_1}}-\bm{\bar{p}_{\xi_2}}\|_2^2+\|\bm{\bar{q}_{\theta_2}}-\bm{\bar{p}_{\xi_1}}\|_2^2\\
  =&4-2(\frac{<\bm{q_{\theta_1}},\bm{p_{\xi_2}}>}{\lVert \bm{q_{\theta_1}} \rVert_2 \lVert \bm{p_{\xi_2}}\rVert_2}+\frac{<\bm{q_{\theta_2}},\bm{p_{\xi_1}}>}{\lVert \bm{q_{\theta_2}} \rVert_2 \lVert \bm{p_{\xi_1}}\rVert_2}).
\end{split}
\label{loss_contrastive}
\end{equation}
We apply an exponential moving average updating strategy to avoid collapse and trivial results during training. We use an optimizer $Opt_2$ to minimize $\mathcal{L}_{cl}$ only with respect to the online network parameters $\bm{\theta}$ as:
   $\bm{\theta} \gets Opt_2(\bm{\theta},\nabla_{\bm{\theta}} \mathcal{L}_{cl})$,
where only $\bm{\theta}$ is updated, and there is a stop-gradient operation for the target network. The target network $\bm{\xi}$ is then updated by the exponential moving average:
    $\bm{\xi} \gets \tau\bm{\xi} +(1-\tau)\bm{\theta}$,  
where $\tau=0.99$ is set as the target delay rate. 
\begin{figure*}[!t]
     \centering
     \begin{subfigure}[b]{0.25\textwidth}
         \centering
         \includegraphics[width=\textwidth]{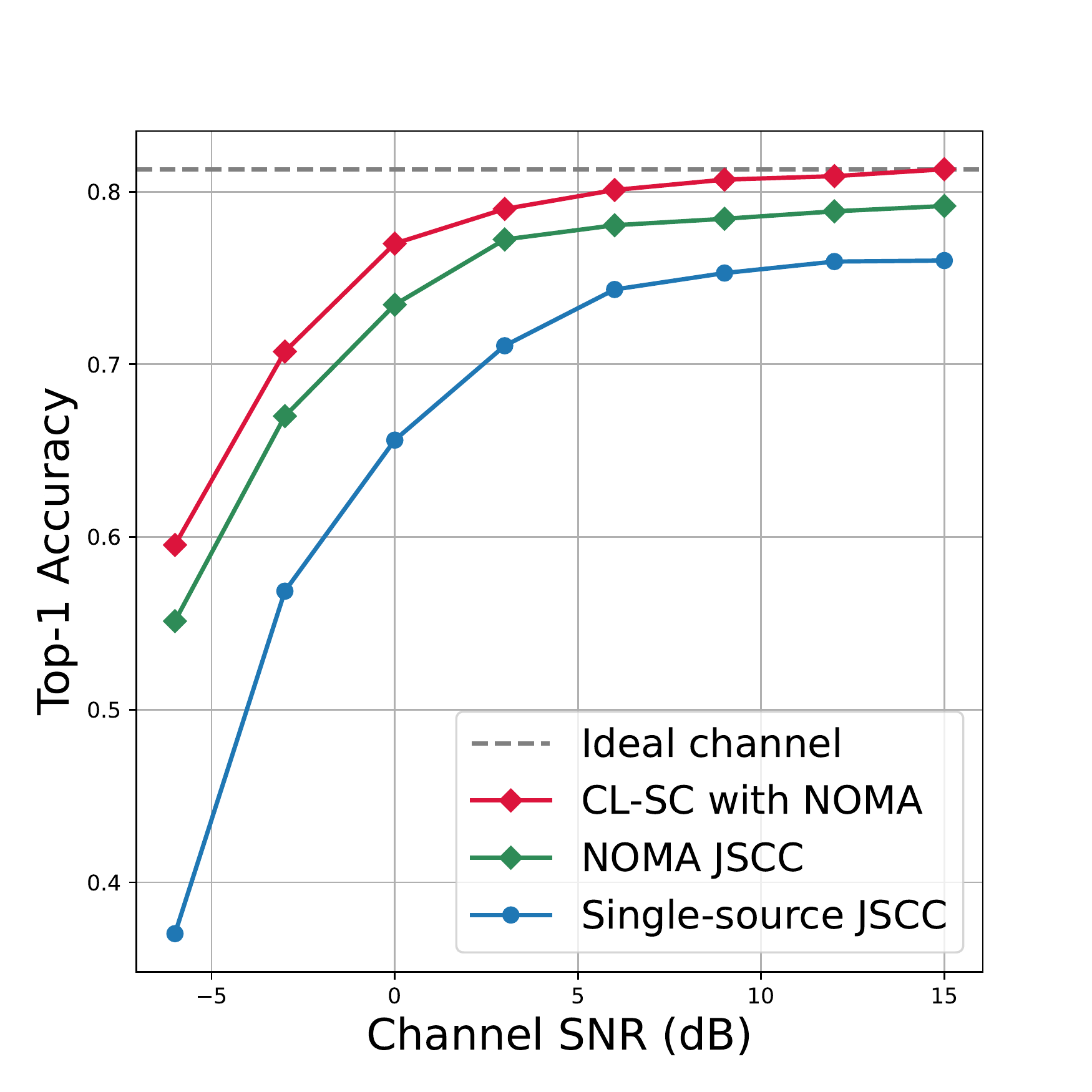}
         \caption{AWGN channel over SNRs}
         \label{fig:noma_awgn_snr}
     \end{subfigure}%
     \begin{subfigure}[b]{0.25\textwidth}
         \centering
         \includegraphics[width=\textwidth]{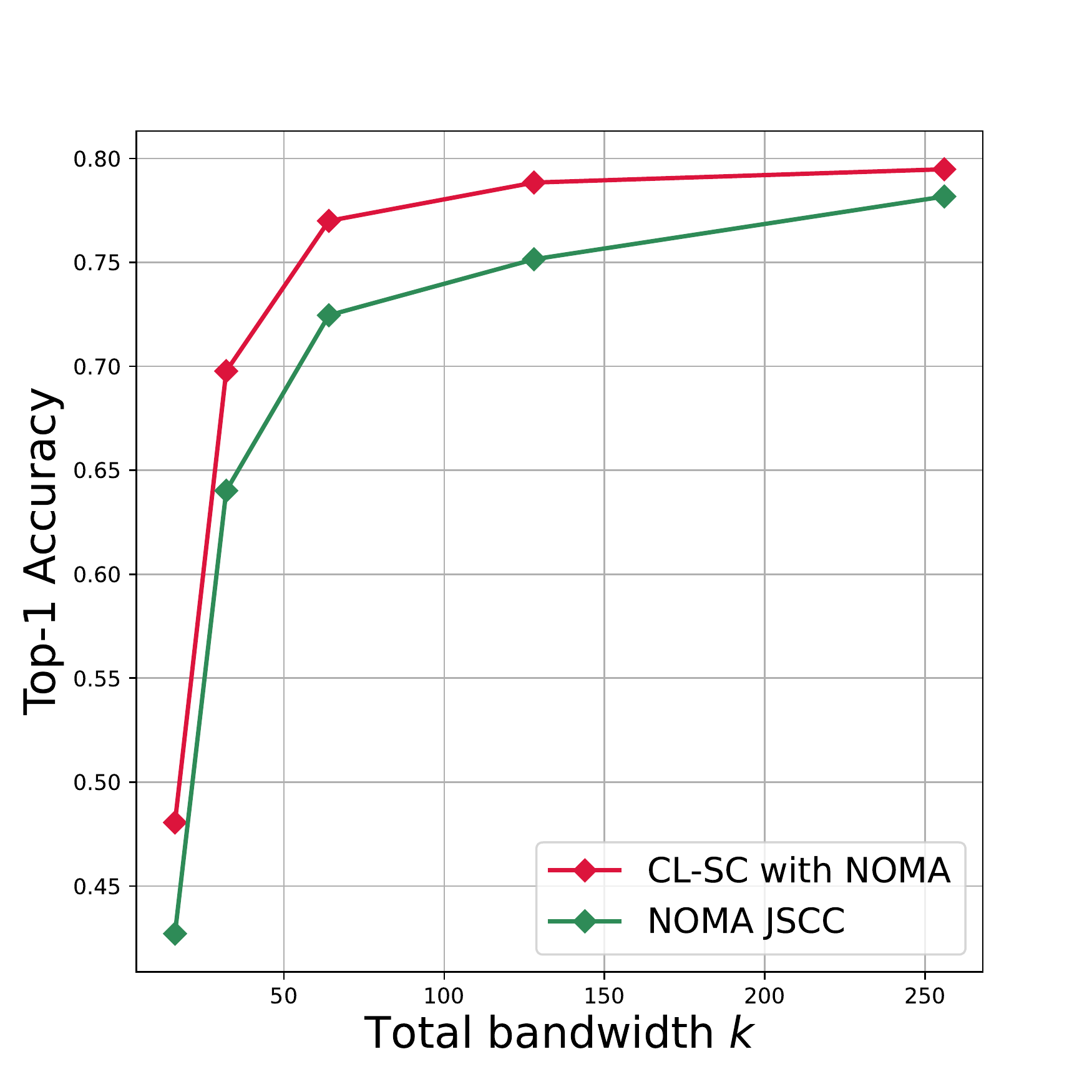}
         \caption{AWGN channel over various $k$}
         \label{fig:noma_awgn_bd}
     \end{subfigure}%
     \begin{subfigure}[b]{0.25\textwidth}
         \centering
         \includegraphics[width=\textwidth]{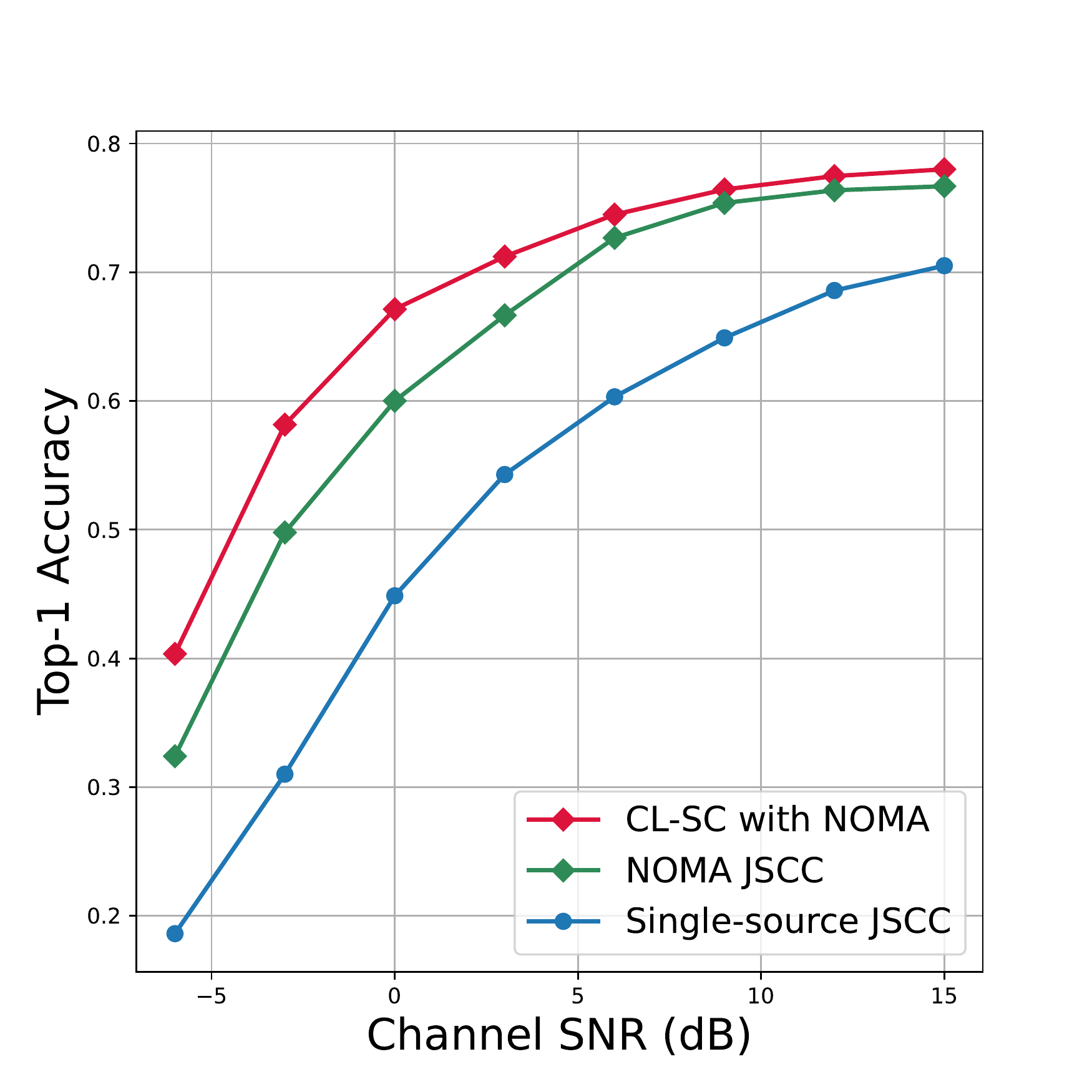}
         \caption{Fading channel over SNRs}
         \label{fig:noma_fading_snr}
     \end{subfigure}%
          \begin{subfigure}[b]{0.25\textwidth}
         \centering
         \includegraphics[width=\textwidth]{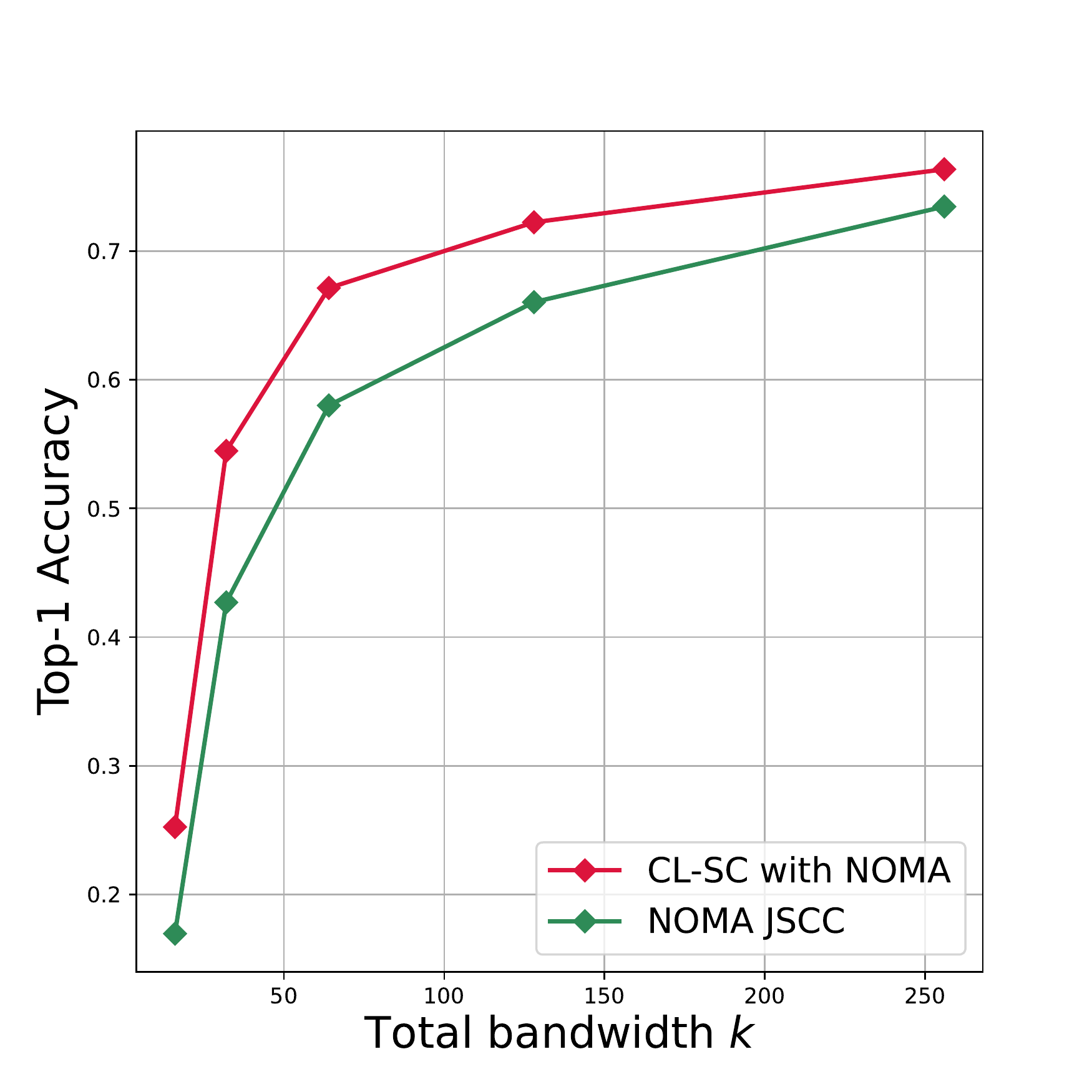}
         \caption{Fading channel over various $k$}
         \label{fig:noma_fading_bd}
     \end{subfigure}%
        \caption{Top-1 retrieval accuracies of the CL-SC scheme over various SNR values and bandwidth $k$, compared with NOMA JSCC\cite{just_draft} and single source JSCC scheme\cite{jscc_ae}.}
        \label{fig:noma_performance}
\end{figure*}

Intuitively, training $\mathcal{L}_{ce}$ should prevent vanishing gradients and improve the performance, while training $\mathcal{L}_{cl}$ should encourage more discriminative features with better representations and semantics. Our numerical results confirm these intuitions.

\section{Training and evaluation}
This section presents numerical experiments to evaluate the performance of our proposed CL-SC scheme under various channel conditions and bandwidths. We consider the NOMA JSCC scheme from \cite{just_draft} and a single-source JSCC scheme as benchmarks. The comparison with the NOMA JSCC scheme from \cite{just_draft} will highlight the gains from our CL module. In the single-source JSCC scheme, a single device transmits without interference. Unless stated otherwise, all models in the figure are trained and tested with the same channel SNR on the Market-1501 \cite{market1501} dataset with two-view data pairs. All experiments are performed with a total channel bandwidth of $k = 64$ symbols, and evaluated with top-1 accuracy\cite{market1501}. To train CL-SC, we sequentially execute a four-step training strategy of modules presented in Fig.~\ref{pipeline}: the pre-training of semantic feature encoders ($T_1$), of JSCC autoencoders ($T_2$), of the CL module ($T_3$), and finally, the end-to-end joint training of all the components ($T_4$). 


We plot the well-trained model performance over various channel SNRs in Figs. \ref{fig:noma_awgn_snr} and \ref{fig:noma_fading_snr} for the AWGN and slow fading channels. We observe that two-source schemes outperform the single-source JSCC scheme at all channel SNRs considered, showing that incorporating multiple views to make a collaborative decision is essential to improve the retrieval accuracy. Compared with the NOMA JSCC benchmark from \cite{just_draft}, we can observe that the proposed CL-SC scheme can significantly improve the top-1 accuracy for all channel SNRs, especially in the low SNR regime, where the CL-SC can improve the top-1 accuracy up to $4.41\%$ and $8.38\%$ for the AWGN and fading channels, respectively. This observation shows that incorporating multi-source correlations with the CL mechanism can significantly improve remote retrieval performance over a wide range of SNRs, especially in bad channel conditions. We also observe that the CL-based scheme can maintain an improvement of $2.12\%$ at high SNRs, which shows the superiority of CL in the multi-view inference problem. Comparing the performance gain over the AWGN channel and fading channels in Fig. \ref{fig:noma_awgn_snr} and Fig. \ref{fig:noma_fading_snr}, respectively, we observe that the CL-SC can improve the performance more significantly in the fading channel. We explain this as the fading channel may result in more distortions and transformations in the transmission disturbance phase, where additional optimization from the CL module is expected to bring more improvements. 


In Fig. \ref{fig:noma_awgn_bd} and Fig.  \ref{fig:noma_fading_bd}, we plot the achieved accuracy over different bandwidth values $k$ in the AWGN and fading channels, respectively, for SNR$=0$dB. As before, the proposed CL-SC scheme outperforms the NOMA JSCC scheme from \cite{just_draft} for all bandwidth values with consistent improvements up to $5.75\%$ and $11.77\%$ for AWGN and fading channels, respectively. In particular, the CL-SC scheme significantly improves the performance in the low bandwidth regimes, such as $k=16, 32, 64,128$. We would also like to highlight that accuracy over $50\%$ can be achieved for a blocklength of slightly higher than $k=16$, which is extremely short for channel coding, and reliable communication at $=0$dB would not be possible. This shows that JSCC is essential for latency-constrained inference at the edge \cite{gunduz2020communicate}.  

\section{Conclusion}
We presented a novel CL-based semantic communication paradigm for a NOMA scheme aimed at a collaborative edge retrieval problem. The proposed method can explicitly explore the signal correlations, and optimize the transmitted features to maximize the retrieval accuracy under given total bandwidth and SNR constraints. Extensive numerical experiments show that exploiting the CL method can significantly improve the retrieval performance for both AWGN and slow fading channels in all the channel conditions and bandwidth values considered, while the improvement is more pronounced in the low SNR and limited bandwidth regimes. The proposed method can be used in any remote communication scenario, where multiple correlated signals are to be exploited for collaborative inference. Extension of this framework to more than two devices is being studied as part of our future work.

\bibliographystyle{IEEEtran}
\bibliography{references}

\end{document}